\documentclass[12pt]{article}
\usepackage{html}
\usepackage{epsf}
\usepackage{graphicx}
\usepackage{amssymb}
\usepackage{hyperref}
\begin{document}
\title{MATTERS OF GRAVITY, The newsletter of the APS Topical Group on 
Gravitation}
\begin{center}
{ \Large {\bf MATTERS OF GRAVITY}}\\ 
\bigskip
\hrule
\medskip
{The newsletter of the Topical Group on Gravitation of the American Physical 
Society}\\
\medskip
{\bf Number 45 \hfill June 2015}
\end{center}
\begin{flushleft}
\tableofcontents
\vfill\eject
\section*{\noindent  Editor\hfill}
David Garfinkle\\
\smallskip
Department of Physics
Oakland University
Rochester, MI 48309\\
Phone: (248) 370-3411\\
Internet: 
\htmladdnormallink{\protect {\tt{garfinkl-at-oakland.edu}}}
{mailto:garfinkl@oakland.edu}\\
WWW: \htmladdnormallink
{\protect {\tt{http://www.oakland.edu/?id=10223\&sid=249\#garfinkle}}}
{http://www.oakland.edu/?id=10223&sid=249\#garfinkle}\\

\section*{\noindent  Associate Editor\hfill}
Greg Comer\\
\smallskip
Department of Physics and Center for Fluids at All Scales,\\
St. Louis University,
St. Louis, MO 63103\\
Phone: (314) 977-8432\\
Internet:
\htmladdnormallink{\protect {\tt{comergl-at-slu.edu}}}
{mailto:comergl@slu.edu}\\
WWW: \htmladdnormallink{\protect {\tt{http://www.slu.edu/colleges/AS/physics/profs/comer.html}}}
{http://www.slu.edu//colleges/AS/physics/profs/comer.html}\\
\bigskip
\hfill ISSN: 1527-3431

\bigskip

DISCLAIMER: The opinions expressed in the articles of this newsletter represent
the views of the authors and are not necessarily the views of APS.
The articles in this newsletter are not peer reviewed.

\begin{rawhtml}
<P>
<BR><HR><P>
\end{rawhtml}
\end{flushleft}
\pagebreak
\section*{Editorial}

Matters of Gravity has adopted a new publication schedule: it will appear in December and June.  The 
purpose of this change is so that a preliminary description of the GGR sessions of each upcoming April APS meeting can  be shown to the GGR membership before the deadline for submission of an abstract for the April meeting.
The next newsletter is due December 2015.  This and all subsequent
issues will be available on the web at
\htmladdnormallink 
{\protect {\tt {https://files.oakland.edu/users/garfinkl/web/mog/}}}
{https://files.oakland.edu/users/garfinkl/web/mog/} 
All issues before number {\bf 28} are available at
\htmladdnormallink {\protect {\tt {http://www.phys.lsu.edu/mog}}}
{http://www.phys.lsu.edu/mog}

Any ideas for topics
that should be covered by the newsletter, should be emailed to me, or 
Greg Comer, or
the relevant correspondent.  Any comments/questions/complaints
about the newsletter should be emailed to me.

A hardcopy of the newsletter is distributed free of charge to the
members of the APS Topical Group on Gravitation upon request (the
default distribution form is via the web) to the secretary of the
Topical Group.  It is considered a lack of etiquette to ask me to mail
you hard copies of the newsletter unless you have exhausted all your
resources to get your copy otherwise.

\hfill David Garfinkle 

\bigbreak

\vspace{-0.8cm}
\parskip=0pt
\section*{Correspondents of Matters of Gravity}
\begin{itemize}
\setlength{\itemsep}{-5pt}
\setlength{\parsep}{0pt}
\item Daniel Holz: Relativistic Astrophysics,
\item Bei-Lok Hu: Quantum Cosmology and Related Topics
\item Veronika Hubeny: String Theory
\item Pedro Marronetti: News from NSF
\item Luis Lehner: Numerical Relativity
\item Jim Isenberg: Mathematical Relativity
\item Katherine Freese: Cosmology
\item Lee Smolin: Quantum Gravity
\item Cliff Will: Confrontation of Theory with Experiment
\item Peter Bender: Space Experiments
\item Jens Gundlach: Laboratory Experiments
\item Warren Johnson: Resonant Mass Gravitational Wave Detectors
\item David Shoemaker: LIGO Project
\item Stan Whitcomb: Gravitational Wave detection
\item Peter Saulson and Jorge Pullin: former editors, correspondents at large.
\end{itemize}
\section*{Topical Group in Gravitation (GGR) Authorities}
Chair: Deirdre Shoemaker; Chair-Elect: 
Laura Cadonati; Vice-Chair: Peter Shawhan. 
Secretary-Treasurer: Thomas Baumgarte; Past Chair:  Beverly Berger;
Members-at-large:
Andrea Lommen, Jocelyn Read, Steven Drasco, Sarah Gossan, Tiffany Summerscales, Duncan Brown, Michele Vallisneri.
Student Member: Jessica McIver.
\parskip=10pt

\vfill
\eject

\vfill\eject

\section*{\centerline
{we hear that \dots}}
\addtocontents{toc}{\protect\medskip}
\addtocontents{toc}{\bf GGR News:}
\addcontentsline{toc}{subsubsection}{
\it we hear that \dots , by David Garfinkle}
\parskip=3pt
\begin{center}
David Garfinkle, Oakland University
\htmladdnormallink{garfinkl-at-oakland.edu}
{mailto:garfinkl@oakland.edu}
\end{center}
Peter Shawhan was elected Vice Chair of GGR; Duncan Brown and Michele Vallisneri were elected members at large of the Executive Committee of GGR. Jessica McIver was elected Student Representative of GGR.  Nicolas Yunes was awarded the IUPAP Young Scientist Prize in General Relativity and Gravitation.

Hearty Congratulations!

\section*{\centerline
{Advanced LIGO in mid-2015}}
\addtocontents{toc}{\protect\medskip}
\addcontentsline{toc}{subsubsection}{
\it Advanced LIGO in mid-2015, by David Shoemaker}
\parskip=3pt
\begin{center}
David Shoemaker, MIT
\htmladdnormallink{dhs-at-mit.edu}
{mailto:dhs@mit.edu}
\end{center}

The Advanced LIGO Project was an NSF-funded replacement of the initial LIGO gravitational-wave detectors, reusing the LIGO Observatory infrastructure.  The NSF granted funds for the effort to fabricate, install, and bring to `lock' three interferometers; the UK, Germany, and Australia also made contributions of designs and hardware. The Project {\it per se} started in 2008 and  the instruments were completed in 2015. The Advanced LIGO (hereafter `aLIGO') Project in fact continues formally until 2017 to allow an optimized procurement and implementation of computing facilities in the scope of the Project. We report here on the completion of the instruments and the state of the commissioning as of the date of writing.

A complementary approach to completion of the instruments at the two observatories was followed, with different aspects of the instruments first available for integrated testing at each site. In this way, lessons learned at each observatory for a certain phase could be transferred to the other observatory, making parallel progress. On the large scale, our plan called for completion at Livingston first, and in May 2014 (just one month after completion of installation) that instrument was brought under full servo-control at the correct operating point and alignment for the optical and mechanical system (`full lock') for multi-hour long intervals. 

At Hanford, there were two instruments to complete: one to be installed and commissioned, and the other put into long-term storage in support of the plan to have it be installed in an infrastructure in India. As a consequence, Hanford planned completion was later than that for Livingston, and the installed instrument at Hanford demonstrated full lock in February 2015. 

Perhaps less interesting to readers of Matters of Gravity, but of great importance to the overall endeavor, a very significant body of documentation on the design, execution, testing, operation, and troubleshooting of the instruments was also completed in the same time frame. With these pre-requisites completed, the instrumental aspect of the project was declared complete by the LIGO Laboratory at the end of March 2015.

Commissioning -- bringing the detectors to an interesting astrophysical sensitivity, and with long locked durations -- is now underway for both instruments. Progress has been surprisingly rapid, and by May 2015 both instruments had demonstrated a sensitivity which could allow the detection of 1.4-solar-mass binary neutron star inspirals to a distance of greater than 50 Mpc (for an SNR of 8, and averaged over directions and polarizations). Detailed noise models for each interferometer help guide the commissioning activity, and at present most of the limitations to sensitivity are well understood and amenable to improvement through tuning and incremental changes. In addition to raw sensitivity, the availability or up-time of the instruments and the stationarity and gaussianity of the strain data are objectives for the commissioning. 

The Virgo Collaboration, of CNRS (France), INFN (Italy), Nikhef (The Netherlands), Hungary and Poland, is also replacing the instrument in the infrastructure in at the European Gravitational-Wave Observatory in Cascina, Italy. Their schedule calls for completion roughly one year later  than that for LIGO, and the installation of the Advanced Virgo instrument is well underway with completion in early 2016 foreseen. A Japanese effort, KAGRA, to build an instrument with unique features of underground construction and cryogenic operation of the optics is also underway. 

The LIGO and Virgo Collaborations established a plan (arXiv:1304.0670 [gr-qc])in early 2013  for the interleaving of commissioning and observing with the network. We are currently `on track' to follow this plan, which calls for a first observing run with the two LIGO instruments to take place in Fall 2015;   the instruments already have sufficient sensitivity for that short three-month run. Further commissioning will follow, and a second, longer observing run, with the Virgo instrument joining LIGO, is planned for Fall 2016. 

The Advanced LIGO Sensitivity once fully commissioned will be roughly ten times better than initial LIGO, and will extend to roughly 10 Hz in contrast to initial LIGO's 40 Hz. Estimates for the detection rates of neutron-star inspirals (arXiv:1003.2480 [astro-ph.HE]) are for roughly 40 detections per year for the ultimate sensitivity (with an order of magnitude uncertainty in rate in either direction). The observing run in Fall 2016 is planned to be at a sensitivity and of a duration such that the there is a reasonable probability of a first detection.  Advanced LIGO, and the other `second generation' detectors, will reach their full sensitivity in the years to follow, allowing both more informative astrophysical interpretations and better pointing for joint electromagnetic and gravitational-wave `multi-messenger' astrophysics.

\vfill\eject

\section*{\centerline
{GGR dominates APS April meeting}}
\addtocontents{toc}{\protect\medskip}
\addcontentsline{toc}{subsubsection}{
\it GGR dominates APS April meeting, by Shane Larson and Deirdre Shoemaker}
\parskip=3pt
\begin{center}
Shane Larson, Northwestern University
\htmladdnormallink{s.larson-at-northwestern.edu}
{mailto:slarson@northwestern.edu}
\end{center}
\begin{center}
Deirdre Shoemaker, Georgia Institute of Technology
\htmladdnormallink{deirdre.shoemaker-at-physics.gatech.edu}
{mailto:deirdre.shoemaker@physics.gatech.edu}
\end{center}

The April APS Meeting in Baltimore this year was an occasion for celebrating the General Relativity Centennial. While the Topical Group in Gravitation always has a strong presence at the April Meeting, the Centennial celebration offered the opportunity to have an impact on the meeting as a whole with an increased number of sessions, more talks, public outreach activities, plenary talks, and exhibits.  We also celebrated our success at achieving 3\% of the APS membership which means that we need to do that one more year and we will be able to apply for division status!

All told, GGR sponsored or co-sponsored 31 Sessions, a total of 200 talks, and 5 plenary talks. In addition, the cover of the meeting program featured a visualization of gravitational radiation by John Baker of GSFC.  We also had several public outreach exhibits including the LIGO traveling exhibit thanks to Marco Cavaglia, the NASA Physics of the Cosmos Booth thanks to Barbara Mattson and Anne Hornschemeier, and the Celebrating Einstein Black (W)hole exhibit thanks to Joey Shapiro Key.

Two special events also provide new opportunities to engage with the gravity community. The first was a public lecture on "Einstein's Legacy: Studying Gravity in War and Peace" by David Kaiser of MIT. Speaking to a packed auditorium, Kaiser described general relativity's spread through the world during the 20th Century, despite the sweep of world history through two world wars and the birth of the Cold War.

\begin{figure}[ht!]
\centering
\includegraphics[width=160mm]{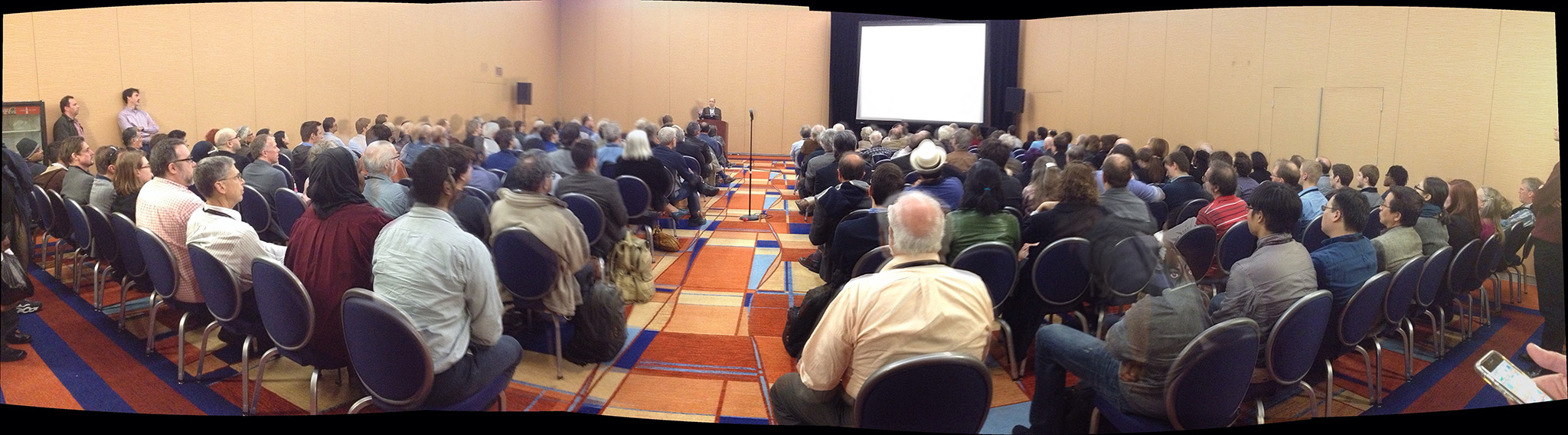}
\end{figure}

\begin{figure}[ht!]
\centering
\includegraphics[width=160mm]{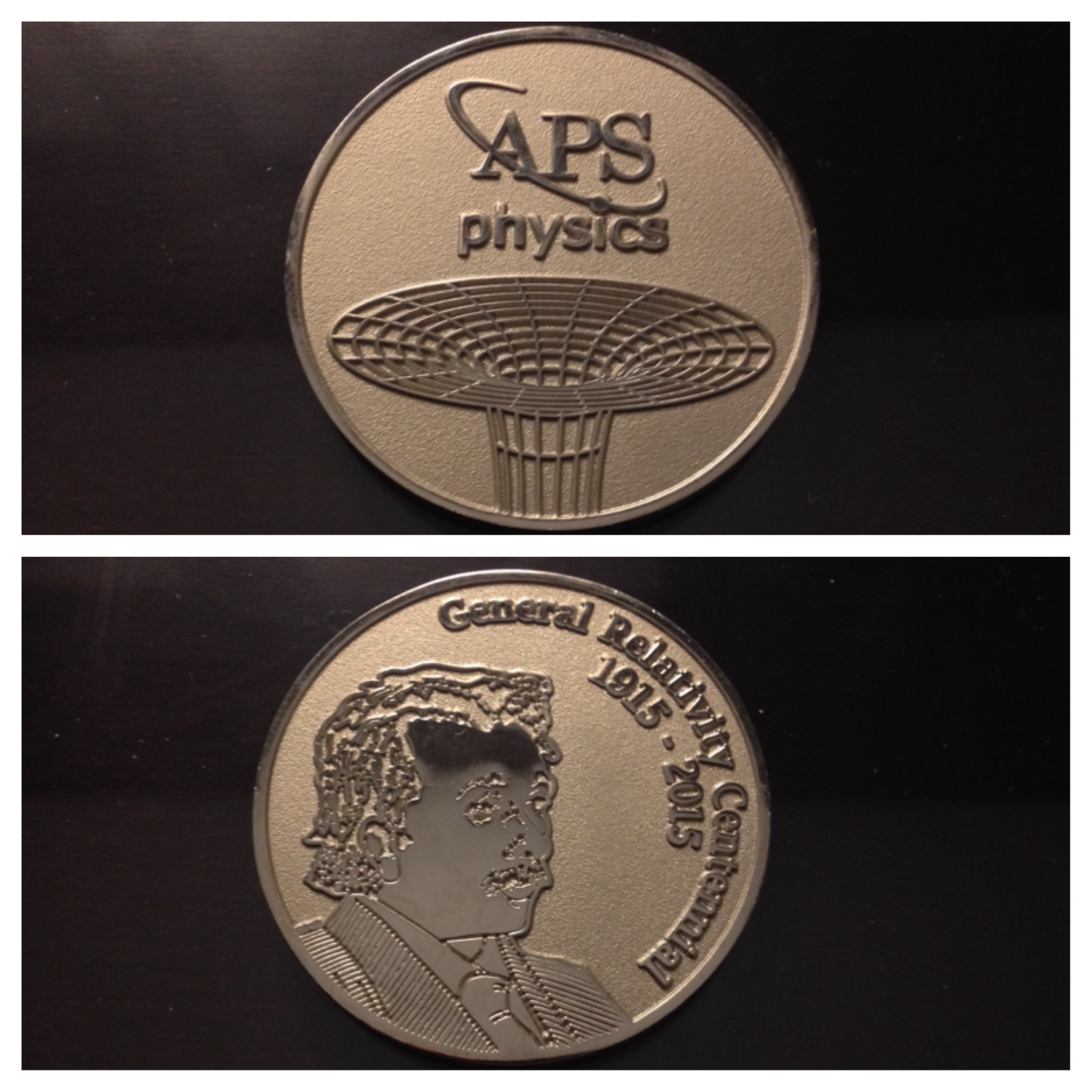}
\end{figure}

The second event was a celebratory banquet, held at Westminster Hall, famously the site Edgar Allan Poe's grave.  The banquet was a festive affair, providing the gravitational physics community at the meeting an opportunity to socialize and celebrate the Centennial together. The after dinner speaker was Richard Price, who reminded us of the dangers arrayed against science and reason in the modern world; he followed this very serious reminder with a (serious?) proposal that we should rise up and form the "$T_{\mu \nu}$ Party" to save the world.  Peter Shawhan also played several clips featuring recordings of Einstein speaking; if you have never heard them, they are highly recommended listening! If you were unable to join us for the celebration, be sure to talk to one of your friends who was there, and see if you can convince them to show you their commemorative medallion. 

All in all, it was an excellent meeting, and a phenomenal opportunity for the gravitational physics community to showcase our science. You can check out more details at the GGR website, http://apsggr.org.  In short: We took over the April meeting; more people were exposed to gravity than ever before! Maybe you should come next year!

\clearpage
\section*{\centerline
{GaryFest}}
\addtocontents{toc}{\protect\medskip}
\addtocontents{toc}{\bf Conference reports:}
\addcontentsline{toc}{subsubsection}{
\it GaryFest, 
by Don Marolf, Veronika Hubeny, and Henriette Elvang}
\parskip=3pt
\begin{center}
Don Marolf, University of California at Santa Barbara
\htmladdnormallink{marolf-at-physics.ucsb.edu}
{mailto:marolf@physics.ucsb.edu}
\end{center}
\begin{center}
Veronika Hubeny, University of Durham
\htmladdnormallink{veronika.hubeny-at-durham.ac.uk}
{mailto:veronika.hubeny@durham.ac.uk}
\end{center}
\begin{center}
Henriette Elvang, University of Michigan
\htmladdnormallink{elvang-at-umich.edu}
{mailto:elvang@umich.edu}
\end{center}
On the occasion of Gary Horowitz's 60th birthday, the two-day conference GaryFest was held at UC Santa Barbara on May 1-2, 2015. Over 100 participants enjoyed the festivities, and contributed to outstanding informal discussions on many subjects (physics and otherwise) that went far beyond the official 16 talks.  The impressive range of the talks, including mathematical relativity, numerical relativity, holography, string theory, and condensed matter physics among others, underscored Gary's versatility and breadth.  

A few photographs from the conference are posted on the main conference web page.  
Slides of the talks can be found by following the blue links at:

\vspace{4mm}
\centerline{\url{http://web.physics.ucsb.edu/~GaryFest/GaryFest\_programme.pdf}}
\vspace{4mm}

\noindent
Video recordings of these lectures\footnote{Except the lectures by Myers and Reall for which the recordings suffered an unfortunate technical glitch.} hosted by the Perimeter Institute's PIRSA will be made available (and linked to the same web page) by the end of June.   For the reader's benefit, we provide a brief overview of the talks below.
\vspace{4mm}

\begin{tabular}{lp{13cm}}
Abhay Ashtekar & Even a tiny positive cosmological constant casts a long shadow \\
& {\it A discussion of physically-useful boundary conditions for studying gravitational radiation in de Sitter space.}
\\
\\
Sean Hartnoll & Entanglement entropy in two dimensional string theory \\
& {\it The $c=1$ matrix model provides a context where one may directly connect entanglement entropy with the corresponding quantity in the bulk dual.}
\\
\\
Thomas Hertog & Holographic Signatures of Cosmological Singularities \\
& {\it A bulk cosmological singularity leads to an intriguing correlation function in the dual field theory.}
\\
\\
Jim Isenberg & On Strong Cosmic Censorship \\
& {\it This classic problem is summarized along with current explorations.}
\\
\\
Luis Lehner & Gary's larger dimensions on Numerical Relativity and his turbulent influence
\\&{\it Luis's personal story about how Gary coerced him into finding the endstate of the Gregory-Laflamme instability, as well as broader issues involving turbulence and non-linear instabilities in empty AdS and black hole spacetimes.}
\\
\\
Juan Maldacena & Cosmological collider physics
\\& {\it At least in principle, CMB non-Gaussianities display quantum mechanical interference phenomena that reflect physics at inflationary energy scales.} 
\\
\\
Robert Myers & Rounding a New Corner with AdS/CMT
\\& {\it The entropy of a region in QFT receives divergent contributions associated with corners -- places where the boundary is not $C^1$.  Some parts of these contributions may be universal. }
\\
\\
Harvey Reall & Causality, hyperbolicity and shock formation in Lovelock theories
\\& {\it The Lovelock equations of motion about certain background are governed by an effective metric that changes signature.  They also produce shocks.  What are the implications?}
\\
\\
Subir Sachdev & Transport in Strange Metals
\\& {\it From CMT to holography and back again: Gravitational and other descriptions of strange metals.}
\\
\\
Jorge Santos & Geons, black holes, and all that Jazz
\\&
{\it What happens to an AdS-Kerr black hole bomb?  It explodes into a hairy black hole having only a single Killing field related to AdS geons.}
\\
\\
Eva Silverstein & The Outside Story: String spreading in the S-matrix and black holes
\\& {\it The effective size of strings grows when they are probed at high frequency.  This growth manifests itself in the S-matrix, and may resolve black hole information puzzles.}
\end{tabular}
\vfill\eject
\begin{tabular}{lp{13cm}}
Andy Strominger & Bardeen-Horowitz Conformal Symmetry in the Sky
\\ 
&{\it Plunging orbits into an extreme black hole -- and the resulting gravitational radiation -- can be computed analytically using the enhanced symmetries of the near-horizon region.}
\\
\\
David Tong & Euclidean Quantum Supergravity
\\& {\it Supersymmetry does not always forbid vacuum energy.  The Kaluza-Klein reduction of 3+1 supergravity on a circle has an effective potential that causes the circle to grow.}
\\
\\
Bob Wald & Black Holes, Thermodynamics, and Lagrangians
\\& {\it Covariant phase space methods give elegant proofs of the first law.  They also link dynamical stability to thermodynamic stability.}
\\
\\
Toby Wiseman & AdS/CFT and the geometry of confinement
\\& {\it The contractable circle in the AdS soliton signals its duality to a confining gauge theory. How can we describe bulk geometries in which more general cycles contract?}
\\
\\
Shing-Tung Yau & Quasilocal mass in general relativity
\\& {\it A concise but pedagogical overview of recent developments regarding quasilocal mass.}
\\
\end{tabular}
\vspace{5mm}

At the banquet, Gary was ``roasted'' by Andy Strominger who expressed his gratitude to Gary for their joint work and long friendship. Many colleagues, past collaborators, students and friends shared their thoughts and stories as well. Outfitted in matching t-shirts, Gary's current students (his 14th, 15th, and 16th) Gavin Hartnett,  Netta Engelhardt, and Eric Mefford presented Gary with a custom-made sweatshirt bearing the name of their ``team": the Santa Barbara Horographers!

At the conference, it was also announced that the Gary T.~and Corinne M.~Horowitz Fellowship has been established to support outstanding beginning PhD students, with preference for students expressing interest in theoretical physics. Created by a group of anonymous donors, the Fellowship honors the contributions of Gary and Corinne to science, UCSB physics, and the global physics community.

It was a terrific conference with interesting science and many wonderful conversations. \\
Happy Birthday Gary!

\end{document}